\documentclass[final,3p,times]{elsarticle}

\usepackage[hidelinks]{hyperref}
\usepackage{lineno}

\usepackage{algpseudocode}
\usepackage{layouts}

\modulolinenumbers[5]

\usepackage{pgf}

\journal{Computers \& Fluids}

\usepackage{amsmath}
\usepackage{placeins}
\usepackage{xcolor}
\usepackage{algorithm}

\usepackage[normalem]{ulem}
\begin{document}

\begin{frontmatter}

\title{Machine Learning for RANS Turbulence Modelling of Variable Property Flows}


\author[mymainaddress]{Rafael Diez\corref{mycorrespondingauthor}}
\cortext[mycorrespondingauthor]{Corresponding author}
\ead{R.G.DiezSanhueza-1@tudelft.nl}
\author[mymainaddress]{Stephan Smit}
\author[mymainaddress]{Jurriaan Peeters}
\author[mymainaddress]{Rene Pecnik}


\address[mymainaddress]{Process and Energy Department, Delft University of Technology, \\Leeghwaterstraat 39, 2628 CB Delft, The Netherlands}

\begin{abstract}
This paper presents a machine learning methodology to improve the predictions of traditional RANS turbulence models in channel flows subject to strong variations in their thermophysical properties. The developed formulation contains several improvements over the existing Field Inversion Machine Learning (FIML) frameworks described in the literature, as well as the derivation of a new modelling technique. We first showcase the use of efficient optimization routines to automatize the process of field inversion in the context of CFD, combined with the use of symbolic algebra solvers to generate sparse-efficient algebraic formulas to comply with the discrete adjoint method. 
The proposed neural network architecture is characterized by the use of an initial layer of logarithmic neurons followed by hyperbolic tangent neurons, which proves numerically stable. The machine learning predictions are then corrected using a novel weighted relaxation factor methodology, that recovers valuable information from otherwise spurious predictions. The study uses the K-fold cross-validation technique, which is beneficial for small datasets. The results show that the machine learning model acts as an excellent non-linear interpolator for DNS cases well-represented in the training set, and that moderate improvement margins are obtained for sparser DNS cases. It is concluded that the developed machine learning methodology  corresponds to a valid alternative to improve RANS turbulence models in flows with strong variations in their thermophysical properties without introducing prior modeling assumptions into the system.


\end{abstract}


\end{frontmatter}



\section{Introduction}
\subsection{Turbulence modelling}


The governing equations of fluid flow have long been established, yet modelling turbulence remains one of the biggest challenges in engineering and physics. 
While it is possible to resolve the smallest scales of turbulent flows using direct numerical simulations (DNS), DNS is still unfeasible for real-world engineering applications. Due to this reason, engineers must rely on RANS turbulence models to describe turbulent flows. However, most of the development of turbulence models has focused on isothermal incompressible fluids. Therefore, these models can be inaccurate when applied to flows with strong variations in their thermophysical properties \cite{OteroReview,HE20084659}, such as supercritical fluids or hypersonic flows. Understanding the behaviour of flows subject to strong property gradients is critical for several engineering applications, such as heat exchangers, supersonic aircraft, turbomachinery, and various applications in the chemical industry~\citep{pecnik2017,Yoo_AR,peeters2022,nemati2015,peeters_jfm_2016}. Even incompressible fluids, such as water, can present large changes in viscosity when subjected to temperature variations.

For incompressible constant-property flows, the governing parameter in the description of turbulent boundary layers is the Reynolds number. For compressible flows, the Mach number and the associated changes in properties become additional parameters that characterize turbulent wall-bounded flows. From past studies, it is known that differences between a supersonic and a constant-property flow can be explained by simply accounting for the mean fluid property variations, as long as the Mach number remains small \citep{smits2006turbulent}. This result is known as Morkovin's hypothesis \citep{morkovin}. 
DNS of compressible channel flows \citep{coleman1995numerical} also suggest that in the near-wall region most of the density and temperature fluctuations are the result of solenoidal 'passive mixing' by turbulence. 
Previous work by \citet{Patel_SLS_2015} has provided a mathematical basis for the use of the semi-local scaling as proposed by \citet{huang1995compressible}. It was concluded that under the limit of small property fluctuations in highly turbulent flows, a change in turbulence is governed by wall-normal gradients of the semi-local Reynolds number, defined as:

\begin{equation}
\label{ReStar_eq}
Re_\tau^{\star} \equiv \frac{\sqrt{\overline{\rho}/{\overline{\rho}_w}}}{\overline{\mu}/{\overline{\mu}_w}}Re_\tau,
\end{equation}
where $\rho$ is the density, $\mu$ dynamic viscosity, the bar denotes Reynolds averaging, the subscript $w$ indicates the value at the wall, and $Re_\tau$ is the friction Reynolds number based on wall quantities and the half channel height, $h$. Thus,  $Re_\tau^{\star}$ provides a scaling parameter which accounts for the influence of variable properties on turbulent flows.

With the semi-local scaling framework and the fact that variable property turbulent flows can be successfully characterized by $Re_\tau^{\star}$, two main developments followed. First, in \citet{patel2016influence}, a velocity transformation was proposed which allows to collapse mean velocity profiles of turbulent channel flows for a range of different density and viscosity distributions. Although following a different approach, this transformation is equivalent to the one proposed by \citet{trettel2016}. Second, this insight has later been used in \citet{pecnik2017} to extend the semi-local scaling framework to derive an alternative form of the turbulent kinetic energy (TKE) equation. 
It was shown that the individual budget terms of this semi-locally scaled (TKE) equation can be characterized by the  semi-local Reynolds number and that effects, such as solenoidal dissipation, pressure work, pressure diffusion and pressure dilatation, are indeed small for the flows investigated. Based on the semi-locally scaled TKE equation, \citet{OteroReview} derived a novel methodology to improve a range of eddy viscosity models. The major difference of the new methodology, compared to conventional turbulence models, is the formulation of the diffusion term in the turbulence scalar equations. 

While these corrections improve the results of RANS turbulence models significantly, they can still be subject to further improvements. Due to these reasons, the present investigation will focus on building ML models to improve the performance of existing RANS turbulence models. 

\subsection{Machine Learning}
In recent years, machine learning has been successfully applied in fluid mechanics and heat transfer due to its inherent ability to learn from complex data, see for instance \citet{CHANG2018815}. While different ML methods are available, deep neural networks have emerged as one of the most promising alternatives to improve turbulence modelling \cite{review_ML4CFD_2020}. These systems are able to approximate complex non-linear functions by using nested layers of non-linear transformations, which can be adapted to the context of every application to optimize the usage of computational resources and to mitigate over-fitting. Different types of neural networks currently hold the state-of-the-art accuracy record in challenging domains, such as computer-vision or natural-language processing \cite{broadreviewML_2015}. During the last decade, one of the main reasons behind the success of deep learning has been the ability of neural networks to both approximate general non-linear functions while providing multiple alternatives to optimize their design.


Significant works in the context of deep learning applied to CFD can be found in the studies of \citet{ling_2016}, who developed deep neural networks to model turbulence with embedded Galilean invariance, or in the work of \citet{Parish_Durai_FIML_07}, where field inversion machine learning (FIML) is proposed in the context of CFD.
Despite the abundance of recent works, significant research is still required regarding the application of ML in the context of CFD, and rich datasets to study turbulence in complex conditions must still be outlined. The future availability of datasets to study turbulence in complex geometries is particularly promising, as this could yield new models with strong applications to industrial and environmental problems.


The methodology for the present study is based on the FIML framework proposed by \citet{Parish_Durai_FIML_07}. This methodology focuses on building corrections for existing RANS turbulence models instead of attempting to rebuild existing knowledge entirely. In the FIML framework, the process of building machine learning models is split into two stages. In the first stage, a data gathering process known as field inversion is performed, where the objective is to identify an ideal set of corrections for the RANS turbulence model under study. Then, in the second stage, a machine learning system is trained in order to replicate the corrections identified. The main advantage of this procedure is that the training process of a neural network is effectively decoupled from the CFD solver, thereby improving the efficiency of the procedure by several orders of magnitude.

For the present work, several modifications are proposed with respect to the study made by \citet{Parish_Durai_FIML_07} and the subsequent publications of \citet{Singh2017_1,Singh2017_2,Singh2017_3}. The modifications considered cover different stages of the problem; such as the optimization methods employed in field inversion, the generation of automatic formulas to compute the gradients of the CFD system, the possibility to automate the process of generating feature groups for the ML system, and novel methods to improve the stability of the FIML methodology while making predictions. 


\section{Fully developed turbulent channel flows}

In this work we consider fully developed turbulent channel flows for which a large number of available DNS studies exist, and for which the time and space averaged conservation equations can be substantially simplified. 

\subsection{DNS database}
The DNS database of turbulent planar channel flows consists of three different sets of simulations. The first set represents variable property low-Mach number channel flows with isothermal walls, heated by a uniform volumetric source to induce an increase of temperature within the channel \cite{pecnik2017,patel2016influence,PatelPRF2017}. Using different constitutive relations for viscosity $\mu$, density $\rho$ and thermal conductivity $\lambda$ as a function of temperature, different DNS cases are used to study the effect of varying local Reynolds and Prandtl number on near wall turbulence. The cases with their respective relations for the transport properties and their corresponding wall-friction velocity based Reynolds number and local Prandtl number are summarized in table~\ref{tableDNScases} (low-Mach number cases). Most of the cases have a friction based Reynolds number at the wall of $Re_\tau$=395. Depending on the distribution of density, viscosity, and conductivity, the semi-local Reynolds number $Re_\tau^{\star}$ and the local Prandtl number are either constant, increasing or decreasing from the walls to the channel center. More details on the cases can be found in Refs.~\cite{patel2016influence, PatelPRF2017, pecnik2017}. 
The second set of DNS consists of high-Mach number compressible channel flow simulations with air modeled as a calorically perfect gas~\cite{trettel2016} (high-Mach number cases). The Mach number ranges from 0.7 to 4 and the corresponding constitutive laws for the transport properties, $Re_\tau$ and Prandtl number $Pr$ are summarized in table~\ref{tableDNScases} as well.  
The third set of simulations contains incompressible channel flows \cite{jimenez2008} (incompressible cases). These cases been added as an additional set to train the FIML framework to account for a large range in Reynolds numbers for incompressible flows.

For all of the variable property DNS cases, it is possible to show that Morkovin's hypothesis applies~\cite{Patel_SLS_2015}. This hypothesis establishes that only the averaged values in molecular properties can be used to characterize the changes in turbulence, and that any higher-order correlations of turbulent fluctuations observed in these properties have a negligible impact in the mean balances  \citep{Patel_SLS_2015,coleman1995numerical}. 
\begin{table}
\begin{center}	
\def\arraystretch{1.2}
\begin{tabular}{ c l c c c c c c c }	
\hline \hline 
{Number} &  {Case ID} & ${\rho}$ & ${\mu}$ & ${\lambda}$ & ${Re_{\tau,w}}$ & {$Pr_w$} & {$Ec_{\tau,w}$} & ${\phi}$ \\ [3pt] \hline 
\multicolumn{9}{c}{Low-Mach number cases \cite{patel2016influence,PatelPRF2017, pecnik2017}} \\ \hline
1 & $CP150$                   & 1         & 1           & 1          & 150 & 1 & 0 & 0     \\ 
2  & CP395                     & 1         & 1           & 1          & 395 & 1 & 0 & 17.55     \\ 
3 & $CP550$                   & 1         & 1           & 1          & 550 & 1 & 0 & 0     \\ 
4  & $CRe_{\tau}^{\star} $       & $T^{-1}$  & $T^{-0.5}$  & 1          & 395 & 1 & 0 & 17.55 \\ 
5  & $SRe_{\tau GL}^{\star}$     & 1         & $T^{1.2}$   & 1          & 395 & 1 & 0 & 18.55 \\ 
6  & $GL$                      & $T^{-1}$  & $T^{0.7}$   & 1          & 395 & 1 & 0 & 17.55 \\ 
7  & $LL1$                     & 1         & $T^{-1}$    & 1          & 150 & 1 & 0 & 29    \\ 
8  & $SRe_{\tau LL}^{\star}$     & $T^{0.6}$ & $T^{-0.75}$ & 1          & 150 & 1 & 0 & 31.5  \\ 
9  & $SRe_{\tau C \nu}^{\star}$  & 1         & $T^{-0.5}$  & 1          & 395 & 1 & 0 & 17.55 \\ 
10  & $C \nu$                   & $T^{-1}$  & $T^{-1}$    & 1          & 395 & 1 & 0 & 16    \\ 
11  & $LL2$                     & 1         & $T^{-1}$    & 1          & 395 & 1 & 0 & 17.55 \\ 
12 & $CRe_{\tau}^{\star}CPr^{\star}$   & $T^{-1}$  & $T^{-0.5}$  & $T^{-0.5}$ & 395 & 1 & 0 & 17.55 \\ 
13 & $GLCPr^{\star}$             & $T^{-1}$  & $T^{0.7}$   & $T^{0.7}$  & 395 & 1 & 0 & 17.55 \\ 
14 & $V\lambda SPr_{LL}^{\star}$ & 1         & 1           & $T^{1}$    & 395 & 1 & 0 & 17.55 \\ 
15 & $CP395_{Pr4}$             & 1         & 1           & 1          & 395 & 4 & 0 & 34    \\ 
16 & $JFM.{CRe_{\tau}^{\star}}$  & $T^{-1}$  & $T^{-0.5}$  & 1          & 395 & 1 & 0 & 95    \\ 
17 & $JFM.{GL}$                & $T^{-1}$  & $T^{0.7}$   & 1          & 950 & 1 & 0 & 75    \\ 
18 & $JFM.{LL}$                & 1         & $T^{-1}$    & 1          & 150 & 1 & 0 & 62    \\ 
\multicolumn{9}{c}{High-Mach number cases \cite{trettel2016}}  \\ \hline
19 & $M0.7R400$ & ~                & ~          & ~          &  437 & ~   & 5.736$\cdot 10^{-4}$ &   \\ 
20 & $M0.7R600$ & ~                & ~          & ~          &  652 & ~   & 5.190$\cdot 10^{-4}$ &   \\ 
21 & $M1.7R200$ & ~                & ~          & ~          &  322 & ~   & 2.804$\cdot 10^{-3}$ &   \\ 
22 & $M1.7R400$ & ~                & ~          & ~          &  663 & ~   & 2.394$\cdot 10^{-3}$ &   \\ 
23 & $M1.7R600$ & $\propto T^{-1}$ & $T^{0.75}$ & $T^{0.75}$ &  972 & 0.7 & 2.135$\cdot 10^{-3}$ & 0 \\ 
24 & $M3.0R200$ & ~                & ~          & ~          &  650 & ~   & 4.751$\cdot 10^{-3}$ &   \\ 
25 & $M3.0R400$ & ~                & ~          & ~          & 1232 & ~   & 4.185$\cdot 10^{-3}$ &   \\ 
26 & $M3.0R600$ & ~                & ~          & ~          & 1876 & ~   & 3.752$\cdot 10^{-3}$ &   \\ 
27 & $M4.0R200$ & ~                & ~          & ~          & 1017 & ~   & 5.574$\cdot 10^{-3}$ &   \\ 
\multicolumn{9}{c}{Incompressible cases \cite{jimenez2008}}  \\ \hline
28 &   $IC.Re180$   & - & - & - & 180  & - & - & - \\ 
29 &   $IC.Re550$   & - & - & - & 550  & - & - & -  \\ 
30 &   $IC.Re950$   & - & - & - & 950  & - & - & -  \\ 
31 &   $IC.Re2000$  & - & - & - & 2000 & - & - & -  \\ 
32 &   $IC.Re4200$  & - & - & - & 4200 & - & - & -  \\ \hline \hline
	\end{tabular}
	\caption{DNS database of turbulent channel flows with variable properties (low-Mach)\citep{patel2016influence,pecnik2017}, with ideal gases at high-Mach numbers \cite{trettel2016}, and with constant properties (incompressible) \cite{jimenez2008}. }	\label{tableDNScases}
\end{center}
\end{table}

\subsection{RANS equations}
To model the turbulent channel flows described above, we use the Reynolds/Favre averaged Navier-Stokes equations. For a fully developed turbulent channel flow, the only in-homogeneous direction of the averaged flow corresponds to the wall-normal coordinate, leading to a set of one-dimensional partial differential equations for the mean momentum, mean energy and any additional transport equations for the turbulence quantities used to close the RANS equations. The Reynolds/Favre averaged streamwise momentum and energy equations for a fully developed turbulent channel flow read 
\begin{align} \label{channel_NSmom}
\frac{\partial }{\partial y} \left[ \left( \frac{\mu}{Re_{\tau,w}} + \mu_t \right) \frac{\partial {u}}{\partial y} \right] &= -1 , \\ 
    \label{channel_NStemp}
	\frac{\partial }{\partial y}
	\left[ \left(\frac{\lambda}{Re_{\tau,w} Pr_w } + \frac{c_p \mu_t}{Pr_t}
	\right) \frac{\partial {T}}{\partial y} \right] &= - Ec_{\tau,w}\left( \frac{\mu}{Re_{\tau,w}} + \mu_t \right) \left(\frac{\partial {u}}{\partial y}\right)^2 -\frac{\phi}{Re_{\tau,w} Pr_w }, 
\end{align}
with the variables $u$ and $T$ referring to the Favre-averaged streamwise velocity and the cross-sectional temperature profiles, respectively. The variables $c_p$, $Pr_t$ and $S_{\phi}$ refer to the isobaric heat capacity, the turbulent Prandtl number, and an arbitrary volumetric heat source term. The coordinates $x$ and $y$ further refer to the streamwise and the wall-normal directions for the channel flow. The wall based friction Reynolds number, the Prandtl number and the friction based Eckert number are defined as 
\begin{equation}
Re_{\tau,w} = \frac{\rho_w u_{\tau,w} h}{\mu_w}, \quad
Pr_w = \frac{\mu_w c_{p,w}}{\lambda_w}, \quad
Ec_{\tau,w} = u^2_\tau/(T_w c_{p,w}) = \left(\gamma-1\right)Ma^2_{\tau,w}, 
\end{equation} 
with $u_{\tau,w} = \sqrt{\tau_w/\rho_w}$ the friction velocity, $h$ the channel half width, $\gamma$ the ratio of specific heats and $Ma_{\tau,w}=u_{\tau,w}/a_w$, where $\tau_w$ is shear stress and $a_w$ the speed of sound at the wall. Given these non-dimensional groups, the non-dimensional density, temperature, viscosity, thermal conductivity are one at the wall, while the non-dimensional isobaric heat capacity is $c_p=1$ in the whole domain.

The mean momentum and energy equations make use of the Boussinesq approximation and the strong Reynolds analogy to model the turbulent shear stress, the turbulent heat transfer, and the turbulent dissipation in the energy equation (first term on the right-hand-side). As such, a turbulent eddy viscosity $\mu_t$ appears in eqs.~(\ref{channel_NSmom}) and (\ref{channel_NStemp}), which is commonly provided by an eddy viscosity model. While many eddy viscosity models exist in literature, in this work we choose a simple $k-\varepsilon$ turbulence model \cite{MKauthors}, which has also been used in our previous studies to model turbulence in variable property turbulent channel flows \cite{OteroReview}. The equations for the turbulent kinetic energy $k$ and turbulent dissipation $\varepsilon$ read 
\begin{align} 
	\label{MK_k_eq}
\underbrace{\mu_t\left(\frac{\partial u}{\partial y}\right)^2}_{P_k} - \underbrace{\rho\ \varepsilon}_{D_k} + \underbrace{\frac{\partial }{\partial y}  \left[\left( {\mu} + \frac{\mu_t}{\sigma_k} \right) \frac{\partial k }{\partial y}  \right]}_{T_k} &= 0, \\
	\label{MK_e_eq}
	 \underbrace{C_{\varepsilon 1}\ P_k\ \frac{\varepsilon}{k}}_{P_\varepsilon}  
	- \underbrace{C_{\varepsilon 2}\ f_{\varepsilon}\ {\rho}\ \frac{\varepsilon^2}{k}}_{D_\varepsilon}
	+  \underbrace{\frac{\partial }{\partial y}  \left[ \left( \mu + \frac{\mu_t}{\sigma_{\varepsilon}} \right) \frac{\partial \varepsilon }{\partial y} \right]}_{T_\varepsilon} &= 0, 
\end{align}
with the supporting damping functions
\begin{equation} 
	\label{MK_fe_eq}
	f_\varepsilon= \left( 1 - \frac{2}{9} e^{-\left(Re_{t}/6\right)^2} \right) \left(1 - e^{-y^{\star}/5}\right)^2, \quad
	f_\mu = \left(1 - e^{-y^{\star}/70} \right) \left(
	1 + \frac{3.45}{\sqrt{Re_{t}}}	\right), 
\end{equation}
and the definition of the turbulent Reynolds number, the semi-locally scaled wall distance and the eddy viscosity, respectively, 
\begin{equation} 
    Re_{t} =  \frac{{\rho}\ k^2}{{\mu}\ \varepsilon}, \quad 
	y^{\star} = y^{+} \sqrt{\frac{ \rho }{ \rho_w }} \frac{ \mu_w }{ \mu }, \quad
	\mu_{t} = C_{\mu}\ f_{\mu}\ {\rho}\ \frac{ k^2}{\varepsilon}.
\end{equation}
The constants take the following values: $C_{\varepsilon 1}=1.4$, $C_{\varepsilon 2}=1.8$, $C_{\mu}=0.09$, $\sigma_{k}=1.4$ and $\sigma_{\varepsilon}=1.3$. Note, the original model uses the wall distance based on viscous wall units $y^+$ in the damping functions. Here, we replaced $y^+$ with $y^{\star}$ to account for the changes in viscous length scales due to changes in density and viscosity close to the wall \cite{OteroReview}.  
A preliminary study revealed that the approximation of $Pr_t \approx 1$ yields accurate temperature profiles for the DNS cases that we will consider. 
The Python and the Matlab source codes to solve the set of RANS equations with the associated boundary conditions can be found on Github~\cite{githubmodel}.

\begin{figure} 
\centering
   \input{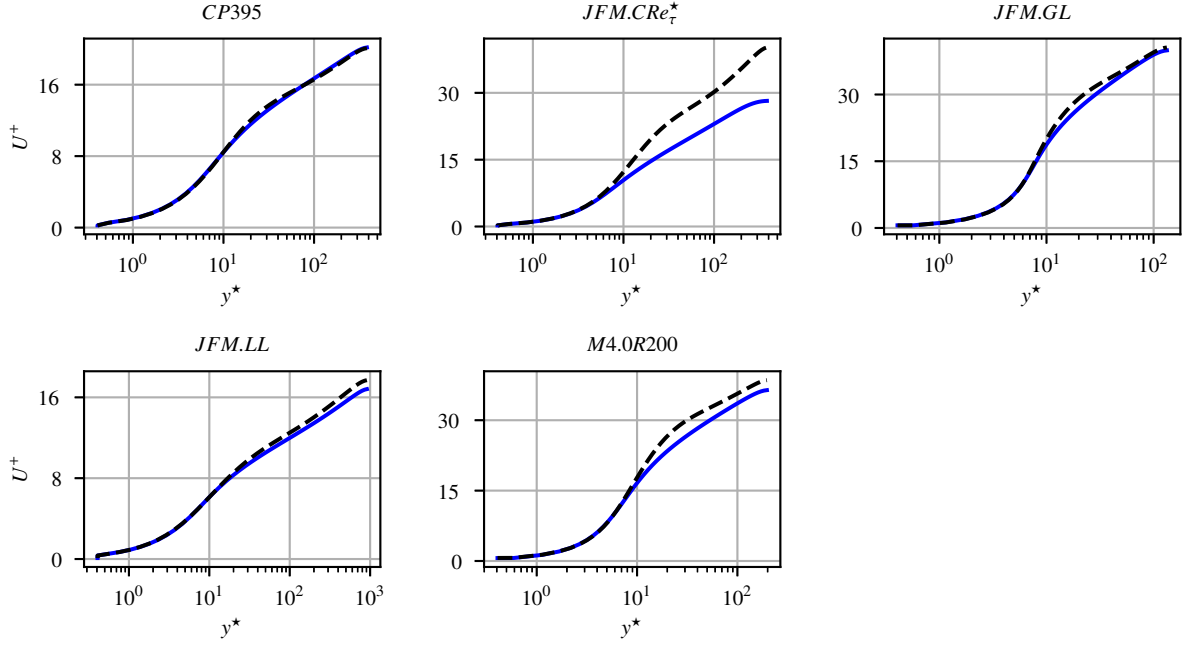}
   \caption{{} Velocity profiles obtained with the original MK turbulence model using the density and viscosity profiles obtained from DNS. The black dashed lines correspond to the DNS data, whereas the blue solid lines correspond to the RANS simulations. }   
  \label{FigMKiniBenchmark}
\end{figure}

The velocity profiles for a few selected cases with the original MK turbulence model are shown in Fig.~\ref{FigMKiniBenchmark}. Large deviations occur in flows subject to strong variable-property gradients. The largest deviations found in such regimes can be found in the DNS case $JFM.CRe_{\tau}^\star$ from Table \ref{tableDNScases}. Here, it can be noted that the maximum error margin reaches a magnitude of 30.4\% at the channel center ($y=H$). Based on these results, it can be noted that the MK turbulence model corresponds to an interesting target for ML optimization, since there exist large deficits to be mitigated. 


\section{Improved field inversion machine learning for variable property turbulence} \label{Subsection_Methodology}
In this section we present an improved methodology of the FIML as proposed by \citet{Parish_Durai_FIML_07}, which is also suitable to account for turbulence in variable-property flows. 



\subsection{Field inversion} \label{Sec_Metho_FI}
In order to minimize the difference between the DNS and the modeled velocity obtained with the RANS approach, the original $k-\varepsilon$ equations are modified by introducing field inversion multipliers $\beta$. The  turbulent kinetic energy $k$ and the turbulent dissipation $\varepsilon$, eqs.~(\ref{MK_k_eq}) and (\ref{MK_e_eq}), can then be written as
\begin{align}
	\label{MK_k_eq_betafied}
    P_k - \beta_k D_k + T_k = 0, \\
    \label{MK_e_eq_betafied}
    P_\varepsilon - \beta_\varepsilon D_\varepsilon + T_\varepsilon = 0. 
\end{align} 
Contrary to \citet{Parish_Durai_FIML_07}, we multiply the dissipation rather than the production terms, in order to adhere to energy conservation, i.e. turbulent kinetic production also appears in the mean kinetic energy equation with an opposite sign \cite{durbin2011statistical}. On the other hand, introducing $\beta_k$ in the $k$-equation can lead to an imbalance of turbulent production and dissipation in the log-layer region. Therefore, we also present results where only $\beta_\varepsilon$ is used to perform the field inversion, since the $\varepsilon$-equation contains the largest amount of empiricism. Another reason to modify the dissipation instead of the production terms is because the production is active in a smaller region of turbulent boundary layers. This implies that field inversion optimizers modifying the dissipation term have a larger capacity to build corrections in regions where other budget terms of RANS turbulence models are still active, such as the diffusion terms. 

It is important to note that field inversion optimizers build corrections, which are ideal with respect to the cost function formulated. As a result, the cost function for the field inversion process must be carefully designed, to minimize not only the differences between the velocity profiles, but also the shape of the  corrections that will be  applied to the turbulence model. A suitable cost function $\mathcal{J}$ is defined as
\begin{equation} 
	\label{MK_Jcost}
	\mathcal{J} = \sum_{i=1}^N I_U \left( \frac{u_i - u^{*}_{i}}{S_U} \right)^2 + I_k \left( \frac{\delta_k}{S_k} \right)^2 + I_{\varepsilon} \left( \frac{\delta_{\varepsilon}}{S_{\varepsilon}} \right)^2, 
\end{equation}
with individual weights $I_U$, $I_k$ and $I_{\varepsilon}$ for each term in the cost function. The first term represents the difference between the RANS velocity profiles ($u$) and the DNS data ($u^*$), whereas the subsequent terms are equivalent to source/sink terms in the turbulence modelling equations. It can easily be shown that $\delta$ is related to $\beta$ as 
\begin{align} 
	\label{MK_Delta_k_eq_ML}
	\delta_k = D_k \left(\beta_k-1\right), \\
	\label{MK_Delta_e_eq_ML}
	\delta_{\varepsilon} =  D_\varepsilon \left(\beta_{\varepsilon}-1\right). 
\end{align}
Finally, $S_U$, $S_k$ and $S_{\varepsilon}$ are used to normalize the variations in the cost function, and they are defined as 
\begin{align} 
	\label{MK_Su_scale}
	S_U &= max(\left|{u^*}\right|),\\
	\label{MK_Sk_scale}
	S_k &= max \left( \left|{P_k}\right| ,\ \left|D_k\right|,\ \left|T_k\right| \right), \\ 
	\label{MK_Se_scale}
	S_\varepsilon &= max \left( \left|{P_\varepsilon}\right| ,\ \left|D_\varepsilon\right|,\ \left|T_\varepsilon\right| \right). 
\end{align}
$\delta_k$ and $\delta_\varepsilon$ are normalized such that the importance factors, $I$, are easier to interpret among all DNS cases considered in this study. As it can be seen in the present formulation, the final field inversion study must include an hyper-parameter optimization analysis for the values of $I_U$, $I_k$ and $I_{\varepsilon}$. The selection method is based on the elbow method \citep{Thorndike1953}, since it was found that each field inversion shows clear inflection points (discussed in detail later).

\subsubsection{Optimization algorithm}
\label{sec:optimizationFI}
To solve the previously defined field inversion problems, we will use gradient-descent (GD) algorithms. 
In general, GD algorithms are preferred over Hessian 
methods to solve complex non-linear optimization problems across different fields. Moreover, for our specific application, it can be shown that the Hessian matrix is  non-invertible at the channel center due to the vanishing gradients near the symmetry plane. 
Accordingly, the $\beta$ multipliers at the channel center have a negligible effect on the solution, and thus their influence on the cost function $\mathcal{J}$ is nearly zero. 
Another favorable property of GD algorithms is that their results yield continuous spatial distributions due to the smoothness of the gradients associated with the turbulence model. Furthermore, GD algorithms tend to leave the $\beta$ multipliers near the channel center at  their initial values ($\beta=1$), since these algorithms do not modify parameters which are not relevant to the cost function $\mathcal{J}$.

\begin{algorithm}
\caption{Modified bold drive method with added momentum to accelerate optimization.}
\label{euclid}
\begin{algorithmic}[1]

\While{$\alpha_{n-1}$ $>$ Threshold}

\State ${m}' \leftarrow c \cdot {m}_{n-1} + \left( 1 - c \right) { \nabla }_{\beta} \mathcal{J}_{n-1}$
\State ${\beta}' \leftarrow {\beta}_{n-1} - \alpha_{n-1} \cdot {m}'$

\If{ $\mathcal{J}\left( {\beta}' \right)<\mathcal{J}_{n-1}$ }
\State ${\beta}_n \leftarrow {\beta}'$
\State ${m}_n \leftarrow {m}'$
\State $\alpha_n \leftarrow k^+  \cdot \alpha_{n-1} $
\Else
\State ${\beta}_n \leftarrow {\beta}_{n-1}$
\State ${m}_n \leftarrow { \nabla }_{\beta} \mathcal{J}_{n-1}$
\State $\alpha_n \leftarrow k^- \cdot \alpha_{n-1}$
\EndIf
\EndWhile
\end{algorithmic}
\end{algorithm}
The GD algorithm used in the present study is based on the traditional bold drive method \citep{BoldDriveBattiti}. However, we also introduced gradient inertia to increase the convergence speed. The final approach for the optimizer is shown in algorithm~\ref{euclid}. In this algorithm, the optimizer starts by taking a traditional step using gradient-descent with added momentum. The values generated for the gradient inertia and the optimization parameters are stored using the auxiliary variables $m’$ and $\beta’$ respectively. If the updated value for the cost function $\mathcal{J}\left( {\beta}' \right)$ is lower than before, the temporary values for $m’$ and $\beta’$ are accepted as the new state of the system. Additionally, the learning rate $\alpha$ is increased according to the expansion ratio $k^+$. This allows the optimizer to dynamically search for a learning rate schedule that maximizes the convergence speed. If divergence is detected ($\mathcal{J}\left( {\beta}' \right)>\mathcal{J}_{n-1}$), the optimizer retains the $\beta$ parameters from the previous iteration ($\beta_{n-1}$), resets the gradient inertia ($m$) to the current Jacobian, and decreases the learning rate according to the ratio $k^-$. These simple steps allow the optimizer to perform a line-search process, seeking optimal values for the learning rate $\alpha$. Gradient inertia must be necessarily removed from the system during the line-search process, since otherwise it cannot be guaranteed that the algorithm will converge to an optimized $\beta$ distribution. 

The recommended values from literature for the parameters $k^+$ and $k^-$ are 1.1 and 0.5, respectively. However, in the present study, we employ an aggressive expansion value of $k^+=1.2$. The constant $c$ corresponds to the gradient inertia hyper-parameter. For this variable, a recommended value of $c=0.9$ can be found across a wide variety of algorithms described in the literature \cite{mom_09used_ref,AdamAlgo}. It was found that the introduction of the gradient inertia decreased the running times by a factor three with respect to the original bold drive method. 
The proposed algorithm allows to fully automatize the process of field inversion, and to subsequently run over 600 optimization cases in total.

\subsubsection{Jacobian matrix calculation}
\label{sec:jacobianFI}
The Jacobian associated with the field inversion process is computed using the discrete adjoint method. In this method, the discretized RANS equations are written as a residual vector $\mathcal{R}(W(\beta),\beta)=0$, that contains one entry per every discretized cell and scalar equation. The variable $W(\beta)$ corresponds to the vector of dicretized physical degrees of freedom present in the RANS equations, such as pressures or velocities. According to the discrete adjoint method, the Jacobian ${ \nabla }_{\beta} \mathcal{J}$ can be calculated as
\begin{equation} 
	\label{DAM_dJ_adj_eq}
	{ \nabla }_{\beta} \mathcal{J} =
	{\Psi}^T \cdot \frac{\partial {\mathcal{R}}}{\partial {\beta}}
	+ \frac{\partial \mathcal{J}}{\partial {\beta}}, 
\end{equation}
where the vector ${\Psi}$ can be obtained from the following system of linear equations
\begin{equation} 
	\label{DAM_Psi_adj_eq}
	\left[ \frac{\partial {\mathcal{R}}}{\partial {W}} \right]^T \cdot
	{\Psi} = - \left[ \frac{\partial \mathcal{J}}{\partial {W}} \right]^T.
\end{equation}
The main advantage of the discrete adjoint method is that only the vector ${\Psi}$ must be calculated from eq. (\ref{DAM_Psi_adj_eq}), whereas a direct calculation method based on chain-rule differentiation would require the computation of the rank 2 sensitivity matrix ${\partial {W}}/{\partial {\beta}}$. Since the latter matrix is orders of magnitude larger than the vector ${\Psi}$, the discrete adjoint method constitutes a better alternative.


In order to generate explicit formulas for all the entries present in the matrices ${\partial {\mathcal{R}}}/{\partial {W}}$ and ${\partial {\mathcal{R}}}/{\partial {\beta}}$, we utilize symbolic algebra packages, such as Sympy~\cite{meurer2017sympy}. As a result, the coefficients of these matrices are described by long arithmetic formulas, which can be inserted into the source code of a function written in any programming language. 
The use of explicit formulas increases the speed of our optimizer as any zero coefficients are immediately cancelled by the algebraic package. Moreover, commonly repeated algebraic sub-terms, such as the eddy viscosity $\mu_t = C_{\mu}\ f_{\mu}\ {\rho}\ { k^2}/{\varepsilon}$, can be replaced by auxiliary variables to avoid redundant calculations. 

\subsection{Neural networks}

In order to complete the FIML methodology, we need to construct a predictive system using neural networks. These networks use hyperbolic tangent neurons in their deeper layers, due to their inherent ability to produce smooth output distributions and since they fulfill the universal approximation theorem \citep{UniApprox1,UniApprox2}. An early reference to the use of hyperbolic tangent neurons in the context of fluid mechanics can be found in the work of \citet{MilanoCFDnn}. In the first layer of the neural network we use logarithmic neurons. These neurons, first proposed by \citet{HinesLogNN}, are capable of automatically discovering input parameter groups, such as the Reynolds number ($Re$), the Prandtl number ($Pr$) or any feature present in boundary layers. The ability to create parameter groups is important, since it is not fully known which features (${X}$) are best suited to model the non-linear relationship ${\delta}= f ({X})$. 
As a result, the introduction of logarithmic neurons into ML systems allows the optimizer to determine which feature groups are optimal; even in the absence of previous modelling knowledge. 

\subsubsection{K-fold validation}

Due to the relatively small size of our database, in combination with the diversity of cases, it proved difficult to split the data between training, cross-validation (CV) and test sets. Picking relevant CV sets that were unbiased by prior turbulence modelling knowledge corresponded to a difficult challenge, since the uniqueness of many DNS samples contained in our database implied that the cases picked for cross-validation could greatly underestimate the error margins found in the test set. As a result, employing CV sets proved to be ineffective.

As a result, the study was performed using the K-fold validation method \cite{Mosteller68}. This method assesses the robustness of machine learning models by picking "K" random training sets, and subsequently evaluating the results with the remaining test set. 
If a large variance is detected, it may indicate that more data is required to train the ML system effectively, or that a different ML architecture is required. The K-fold validation method corresponds to one of the best alternatives available to assess the performance of ML systems trained with small datasets \cite{kfold_case_study}.  
 
The test sets for the different K-fold combinations are listed in table~\ref{tableKfoldTrials}. The DNS cases are picked randomly, except for the K-fold set (K-1), which corresponds to a hand-picked selection of challenging test cases yielding large modelling errors when using the MK turbulence model. 
Among the test cases of the K-fold set (K-1), DNS cases are included that are subject to extreme conditions; such as the cases $JFM.CRe^{\star}_{\tau}$ and $M4.0R200$. As a result, the K-fold set (K-1) represents a scenario where challenging predictions are required, despite the absence of adequate training samples. The incompressible DNS cases extracted from the work of \citet{jimenez2008} are added to the test sets of the K-fold validation trials (K-2 to K-10) in order to assess the response of the ML system for different Reynolds numbers. However, these DNS cases were omitted for the K-fold set (K-1), since the intent of this validation run was to test the ML system only under challenging variable-property flow cases.
\begin{table} 
	\begin{center}	
	\def\arraystretch{1.2}
	\begin{tabular}{ l l l l l l l }	
	\hline \hline
K-1 & $JFM.CRe^{\star}_{\tau}$ & $C \nu$ & $CRe^{\star}_{\tau}CPr^{\star}$ & $M3.0R200$ & $M4.0R200$ \\   \hline 
K-2 & $JFM.CRe^{\star}_{\tau}$ & $SRe^{\star}_{\tau LL}$ & $GLCPr^{\star}$ & $M1.7R200$ & $M4.0R200$ & $IC.Re950$ \\      \hline 
K-3 & $JFM.CRe^{\star}_{\tau}$ & $JFM.GL$ & $JFM.LL$ & $M3.0R600$ & $M4.0R200$ & $IC.Re550$ \\        \hline 
K-4 & $JFM.CRe^{\star}_{\tau}$ & $LL2$ & $CP150$ & $V \lambda SPr^{\star}_{LL}$ & $M3.0R200$ & $IC.Re950$ \\        \hline 
K-5 & $JFM.LL$ & $LL1$ & $C \nu$ & $CP395_{Pr 4}$ & $M0.7R400$ & $IC.Re550$ \\  \hline 
K-6 & $JFM.GL$ & $C \nu$ & $CRe^{\star}_{\tau}CPr^{\star}$ & $M0.7R400$ & $M4.0R200$ & $IC.Re2000$ \\       \hline 
K-7 & $JFM.CRe^{\star}_{\tau}$ & $GL$ & $CRe^{\star}_{\tau}CPr^{\star}$ & $GLCPr^{\star}$ & $M1.7R200$ & $IC.Re550$ \\  \hline 
K-8 & $JFM.GL$ & $SRe^{\star}_{\tau GL}$ & $SRe^{\star}_{\tau LL}$ & $V \lambda SPr^{\star}_{LL}$ & $M1.7R400$ & $IC.Re4200$ \\   \hline 
K-9 & $JFM.LL$ & $CP150$ & $GLCPr^{\star}$ & $CP395_{Pr 4}$ & $M1.7R600$ & $IC.Re180$ \\      \hline 
K-10 & $JFM.CRe^{\star}_{\tau}$ & $SRe^{\star}_{\tau C \nu}$ & $CRe^{\star}_{\tau}CPr^{\star}$ & $CP395_{Pr 4}$ & $M0.7R400$ & $IC.Re950$ \\ 	\hline \hline
	\end{tabular}
	\caption{Test cases considered for the implementation of the K-fold validation method for the machine learning study. }	\label{tableKfoldTrials}
		\end{center}
\end{table}
 
The selection procedure to determine the final machine model for the study is based on finding the smallest neural network architecture which is capable of fitting the training data available for the K-fold combination (K-1) listed in Table \ref{tableKfoldTrials}. According to the principles of the elbow method, the smallest system which can fit the training data is less likely to produce over-fitting than larger machine learning models. The K-fold set (K-1) is chosen, since this combination represents a realistic scenario where a selection of the most challenging CFD cases remain hidden from the training data. In summary, the final neural network architecture is not pre-conditioned to perform well under the most complex test conditions available. 

\subsubsection{Weighted relaxation factor}
\label{sec:relaxmethod}
\begin{figure}
\centering
    \input{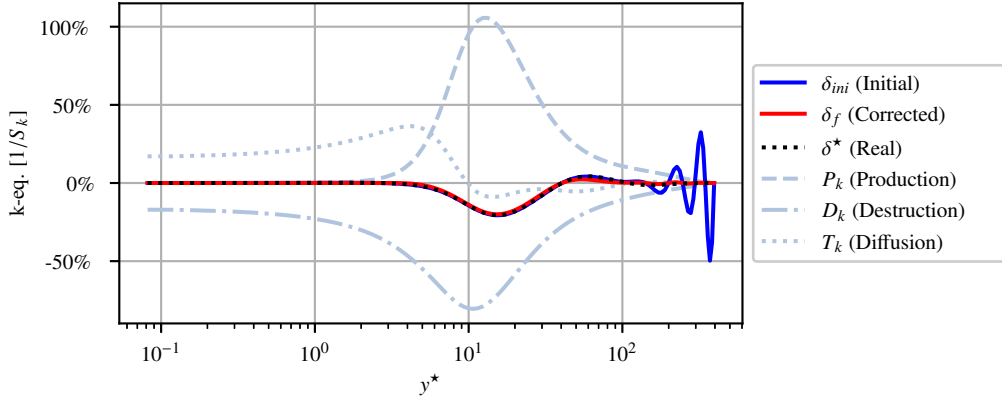}
    \centering
  	\caption{Turbulence budgets for the optimized k-equation of the MK turbulence model after applying the independent set of ${\delta_k}$ corrections obtained during the field inversion study for the DNS case $JFM.CRe_{\tau}^{\star}$. The initial machine learning predictions, shown in blue, contain the ficticious perturbation term: $\Delta=0.6\ y^3\ sin \left( 8 \pi y \right)$. The red line $\delta_{ML}$ corresponds to the corrections obtained after applying the weighted relaxation factor methodology.}
  	\label{Fig_First_Alpha}
\end{figure}

Fig.~\ref{Fig_First_Alpha} shows a comparison between the ${\delta_{k}}$ predictions, made by a fictitious neural network ($\delta_{ini}$), and the ground-truth labels ${\delta^*}$ obtained through field inversion for the DNS case $JFM.CRe^{\star}_{\tau}$. The values of $\delta_{ini}$ in Fig.~\ref{Fig_First_Alpha} contain the explicit perturbation term $\Delta=0.6\ y^3\ sin \left( 8 \pi y \right)$, which was added to test the robustness of the present methodology. The corrections present in the distribution ${\delta_{ini}}$ are physically plausible up to  $y^{\star}<100$, while spurious 
oscillatory corrections exist at the channel center. These oscillations also produce unstable behavior in CFD solvers, since all budget terms in the RANS equations are inactive near the symmetry plane at the channel center. To mitigate the previous issues, we introduce a weighted relaxation factor methodology, $\alpha$, which is defined as the fraction of the original ML corrections $\delta_{ini}$ to keep.

The derivation of the weighted relaxation factor starts by noting that the magnitude of the final corrections that will be applied to the RANS model, ${\delta_f}$, only corresponds to a fraction, $\alpha$, of the original corrections predicted by a neural network, ${\delta_{ini}}$. This relation can be stated as
\begin{equation} 
	\label{IRF_Fixed_mag1}
	\left|\left|{\delta_{f}}\right|\right| = \alpha\ \left|\left|{\delta_{ini}}\right|\right|,
\end{equation}
or alternatively, 
\begin{equation} 
	\label{IRF_Fixed_mag2}
	\mathcal{J}_{\delta} = \sum_{i=1}^N \delta_{f,i}^2  = \sum_{i=1}^N \left(\alpha\ \delta_{ini,i}\right)^2 \ (= constant).
\end{equation}
In order to assess the true compatibility of the final ${\delta_{f}}$ corrections with a given RANS turbulence model, we express $\delta_{f}$ in eq.~(\ref{IRF_Fixed_mag2}) as ${\beta}$ times the production term ${P}$, namely 
\begin{equation} 
	\label{IRF_Fixed_mag3}
	{\delta_{f}} = {\beta} {P} 
\end{equation}
in the corresponding turbulence modelling equations. 
They key idea to build a robust methodology is to recognize that spurious machine learning corrections, such as ${\delta_{ini}}$ in Fig.~\ref{Fig_First_Alpha}, create large oscillations in the $\beta$ multipliers defined by eq. (\ref{IRF_Fixed_mag3}). As a result, the introduction of a L2-regularization hyper-parameter, $\lambda$, for these ${\beta}$ multipliers would immediately penalize the presence of large oscillations in regions where RANS turbulence models are inactive. The cost function associated with  this problem is 
\begin{equation} 
	\label{IRF_Fixed_mag5}
		\begin{split}
	\mathcal{J}_{\beta} & = \sum_{i=1}^N \left(\delta_{f,i}-\delta_{ini,i}\right)^2
	+ \lambda \beta_i^2 \\
	 &=  \sum_{i=1}^N \left(P_i \beta_{i} -\delta_{ini,i}\right)^2
	+ \lambda \beta_i^2.
		\end{split}
\end{equation}

Eq. (\ref{IRF_Fixed_mag5}) states that the final ${\beta}$ multipliers must produce the greatest degree of similarity between ${\delta_{f}}$ and ${\delta_{ini}}$, while minimizing the magnitude of $\left|\left|{\beta^2}\right|\right|$ according to a regularization hyper-parameter $\lambda$. In order to minimize the cost function defined in eq. (\ref{IRF_Fixed_mag5}), its Jacobian can be forced to form a null vector:
\begin{equation} 
	\label{IRF_Fixed_mag6}
	{ \nabla }_{\beta} \mathcal{J}_{\beta} = {0}.
\end{equation}
Replacing eq. (\ref{IRF_Fixed_mag5}) into the previous condition, yields the following element-wise array equation:
\begin{equation} 
	\label{IRF_Fixed_mag7}
	{P} \left({P} {\beta}-{\delta_{ini}}\right)+\lambda {\beta}   =0.
\end{equation}
Re-arranging the terms of eq. (\ref{IRF_Fixed_mag7}) further reveals that 
\begin{equation} 
	\label{IRF_Fixed_mag9}
	{\beta}  =\frac{{\delta_{ini}}\ {P}}{  \lambda + {P}^2 }.
\end{equation}
Note, eq.~(\ref{IRF_Fixed_mag9}) is evaluated element-wise. Replacing eq.~(\ref{IRF_Fixed_mag9}) back into eq.~(\ref{IRF_Fixed_mag3}) gives a direct residual equation for $\lambda$ 
\begin{equation} 
	\label{IRF_Fixed_mag10}
	\mathcal{R}_{\lambda} = \sum_{i=1}^N \left(\delta_{ini,i}\ \frac{{ P_i^2}}{\lambda + {P_i^2}} \right)^2  - \sum_{i=1}^N \left(\alpha\ \delta_{ini,i}\right)^2  =0.
\end{equation}
Since eq. (\ref{IRF_Fixed_mag10}) only contains one unknown ($\lambda$), a simple root-finding algorithm can be used to solve this optimization problem, such as the Newton--Raphson method. 
For reference, the gradient of the previous residual equation ($\mathcal{R}_{\lambda} $) is given by the following formula:
\begin{equation} 
	\label{IRF_Fixed_mag12}
 { \nabla }_{\lambda} \mathcal{R}_{\lambda} =  -2 \sum_{i=1}^N  \frac{\left(  \delta_{ini,i} P_i^2 \right)^2}{\left(\lambda + {P_i^2}\right)^3}.
\end{equation}
After obtaining the regularization hyper-parameter, $\lambda$, the final ML corrections (${\delta_{f}}$) are given by:
\begin{equation} 
	\label{IRF_Fixed_mag13}
	{\delta_{f}} = \frac{{\delta_{ini}}\ {P}^2}{  \lambda + {P}^2 }.
\end{equation}
Eqs. (\ref{IRF_Fixed_mag10}-\ref{IRF_Fixed_mag13}) constitute the only required components to implement our weighted relaxation factor methodology in a computer environment. The results depicted in Fig.~\ref{Fig_First_Alpha}, show how the relaxation factor removes 40\% of the worst part of the initial corrections, ${\delta_{ini}}$, based on the budget for the production term of the $k$-equation. The final distribution obtained, effectively resembles the ground-truth labels, ${\delta^*}$, which were hidden from the system. 

Finally, it is important to note that the current weighted relaxation factor methodology is applicable to either \mbox{1D}, \mbox{2D} or \mbox{3D} RANS turbulence models. The systems of equations obtained in each case are identical, since the formulation employed does not impose constraints on the spatial distribution of the data points considered.

\subsection{Final framework}

\begin{figure}
\centering
    \input{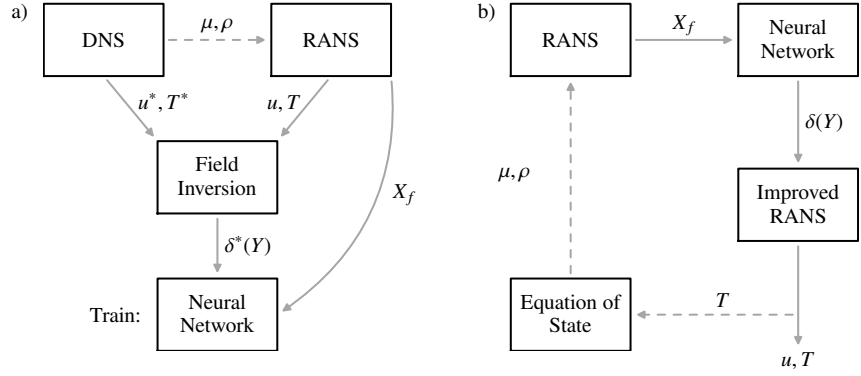}
    \centering
  	\caption{Final framework established for the FIML methodology. The dashed lines indicate the additional steps which are necessary to handle the presence of variable-property flows, with respect to the original scheme proposed by \citet{Parish_Durai_FIML_07}. The diagram on the left (a) presents the methodology employed to obtain field inversion corrections and train deep learning systems, whereas the scheme on the right (b) corresponds to the feedback loop used to used to perform predictions at runtime.}
  	\label{Fig_C4FIMLframework}
\end{figure}

The final machine learning framework developed can be found in Fig.~\ref{Fig_C4FIMLframework}, which is split into two stages. In the first stage, shown in Fig.~(\ref{Fig_C4FIMLframework}.a), a baseline RANS turbulence model is compared with existing DNS data in order to produce $\delta_k$ field inversion corrections. These corrections are subsequently used as ground-truth labels to train a deep learning system, which is able to replicate the trends observed based on a stack of relevant features $X_f$. One of the main differences between the framework described in Fig.~(\ref{Fig_C4FIMLframework}.a) and the original approach proposed by Parish \& Duraisamy \cite{Parish_Durai_FIML_07} is that our field inversion corrections $\delta_k$ are subject to L2-regularization based on their absolute magnitude as a fourth budget-term in the RANS equations, instead of their values as relative $\beta$ multipliers with respect to existing RANS terms. This enables the field inversion optimizer to build corrections that follow patterns which are not captured by baseline RANS models. Additionally, the framework described in Fig.~\ref{Fig_C4FIMLframework}(a) has been adapted to account for the changes observed in flows subject to strong variations in their thermophysical properties, namely, $\rho$, $\mu$ and $\lambda$. The approach chosen effectively decouples the analysis of the RANS momentum equations from the energy equation or any associated equations-of-state for fluids. This is achieved by passing the DNS profiles for the molecular properties of the fluids to the baseline RANS turbulence models during the field inversion process. However, one challenge introduced by this procedure is that the stack of features $X_f$ used to predict the field inversion corrections $\delta_k$ must be based on accurate estimations of the profiles for the molecular properties of fluids. This challenge was solved by using the machine learning framework presented in Fig.~\ref{Fig_C4FIMLframework}(b), which corresponds to the system employed to create predictions for unknown CFD cases. This algorithm is based on an iterative feedback loop, where the improved velocity profiles generated by the predicted $\delta_{ML}$ corrections are passed back to the viscous heating term of the energy equation in order to generate better estimates for the profiles of the molecular properties $\mu$ and $\rho$. The algorithm is initialized by using the profiles generated by the baseline RANS model without $\delta$ corrections, and it iterates until convergence is achieved. The weighted relaxation factor methodology described in section \ref{sec:relaxmethod} is applied immediately once neural network predictions are obtained.

\begin{figure}
\centering
    \input{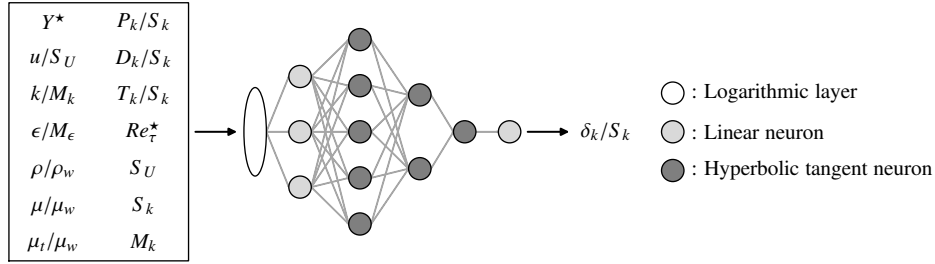}
    \centering
  	\caption{Neural network architecture created to predict the field inversion corrections (${\delta_k}$) required by the MK turbulence model.}
  	\label{Fig_NNDeltaK}
\end{figure}

The final neural network architecture is depicted in Fig.~\ref{Fig_NNDeltaK}. Here, it can be seen that only three logarithmic neurons are used in the initial layer. The initial stack of features, presented in Fig.~\ref{Fig_NNDeltaK}, corresponds to different physical quantities that may be considered by the NN. The previous quantities are intended to be computed based on the initial turbulence budgets found in the uncorrected RANS equations. The sub-scales $M_k$ and $M_\varepsilon$ correspond to references used to normalize the scalar fields $k$ and $\varepsilon$ based on the magnitude of the destruction terms in the RANS equations:
\begin{equation} 
	\label{NN_eq_Me}
	M_{\varepsilon} = \frac{S_{k}}{\rho_w }, \quad 
	M_k = \frac{\rho_w {M_{\varepsilon}^2}}{S_{\varepsilon}}.
\end{equation}
	

\FloatBarrier
\section{Field inversion results}
This section will describe the results of the FIML study for the MK turbulence model. First, the different hyper-parameter combinations for the field inversion study will be analyzed. Then, the observed trends in the final machine learning predictions will be presented, followed by a brief discussion of the results.

\begin{figure}
\centering
    \input{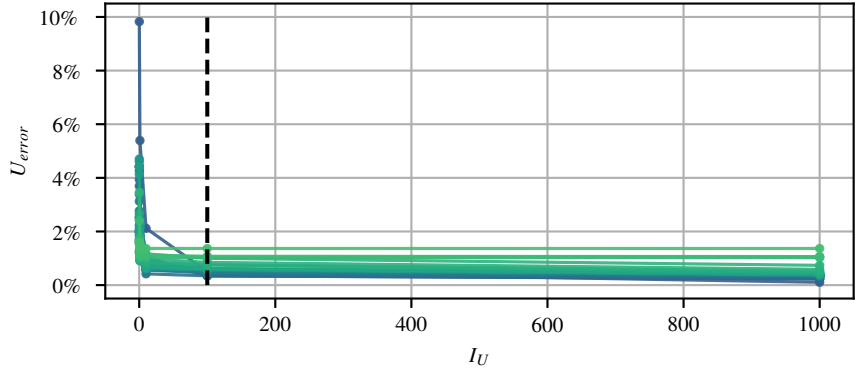}
    \centering
  	\caption{Application of the elbow method to determine the magnitude of $I_U$ during the initial field inversion study for the MK turbulence model ($I_k=I_{\varepsilon}=1$). The black dashed line represents the position where $I_U=100$.}
  	\label{Fig_elbowIu_Ik1Ie1}
\end{figure}

The field inversion study of the MK turbulence model focuses on determining the values of the hyper-parameters $I_U$, $I_k$ and $I_\varepsilon$ in eq.~(\ref{MK_Jcost}). In the first combination, an equal importance is assigned to the corrections used in each scalar equation ($k$ and $\varepsilon$) by setting  $I_k=I_\varepsilon=1$. The value of $I_U$ was calibrated by applying the elbow method to the system. The results obtained can be found in Fig.~\ref{Fig_elbowIu_Ik1Ie1}, where it can be seen that clear inflection points exist for each case. For any subsequent ML analysis, it is possible to either choose the $I_U$ values located at the inflection point of each DNS case, or to pick a common value of $I_U$ for all cases. It was decided to pick a common value of $I_U=100$ for all DNS cases, since this creates smooth trends across the whole dataset. Additionally, selecting a unique value for $I_U$ can simplify the creation of deep learning models, since all the target $\delta^*$ corrections correspond to the solution of a single optimization problem. If different $I_U$ values were picked for each DNS case, additional training data might be required to allow the deep learning system to approximate the selection criterion employed.

\begin{figure}
\centering
    \input{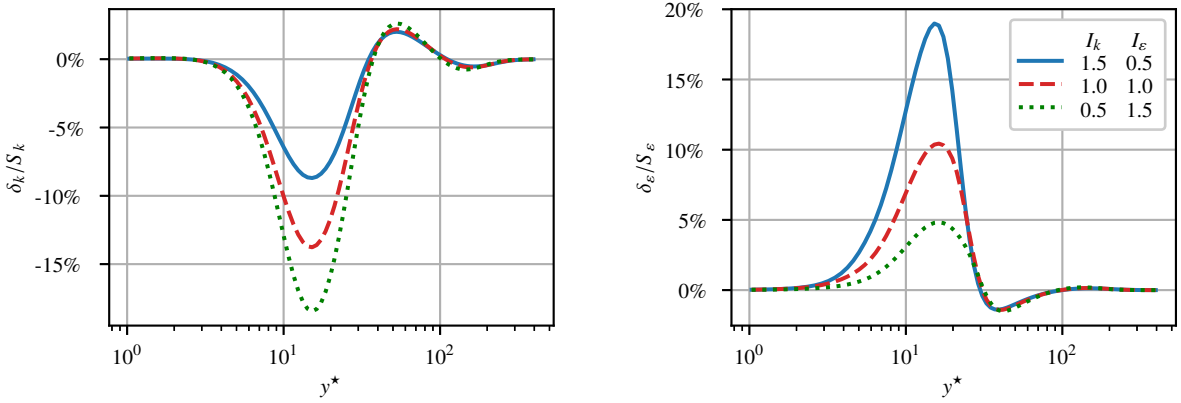}
    \centering
  	\caption{Effect of the cross-interactions between $I_k$ and $I_{\varepsilon}$ for $I_U=100$ during the field inversion study of the MK turbulence model for the DNS case $JFM.{CRe_{\tau}^\star}$.}
  	\label{Fig_Sensi_Crets_IkIe}
\end{figure}

The second stage of the hyper-parameter optimization study consists in analyzing the effect of the individual values of $I_k$ and $I_\varepsilon$ in the field inversion results. The effects of varying these hyper-parameters are depicted in Fig.~\ref{Fig_Sensi_Crets_IkIe} for the DNS case $CRe^{\star}_{\tau}$, which corresponds to the case with the highest modelling errors according to the MK turbulence model. The results show that varying the values of $I_k$ and $I_\varepsilon$ has a minor influence in the shape of the corrections, since only the magnitude of the peaks tends to change. After a detailed analysis, it can be proven that the trends observed in Fig.~\ref{Fig_Sensi_Crets_IkIe} are also present across different DNS cases, and that further tuning the values of $I_k$ and $I_\varepsilon$ is not relevant in the context of the present study. 

\begin{figure}
\centering
    \input{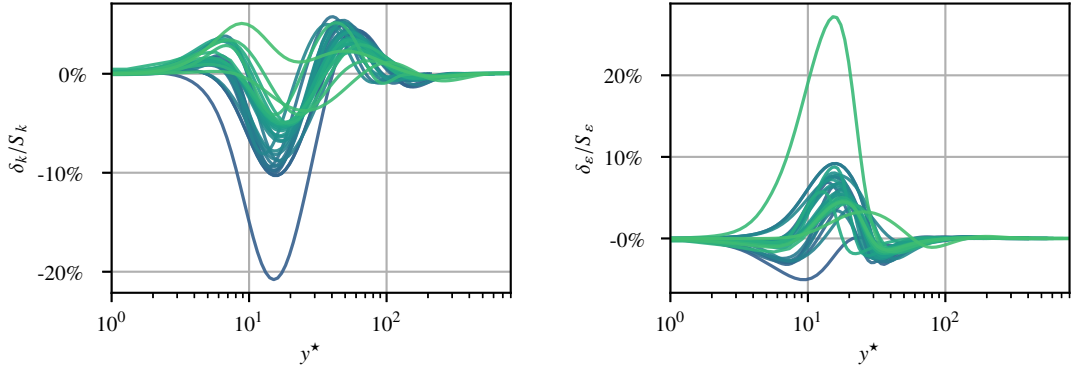}
    \centering
  	\caption{Independent set of field inversion corrections ${\delta_k}$ and ${\delta_{\varepsilon}}$ obtained for the MK turbulence model while employing $I_U=100$.}
  	\label{Fig_DeltaIu100_DeltaKIe0}
\end{figure}

After analyzing the effect of the different hyper-parameters found in the field inversion formulation, it was decided to study the effect of building independent sets of $\delta_k$ and $\delta_{\varepsilon}$ corrections. Employing a unique set of corrections can simplify the subsequent ML study, since the need to produce two-dimensional output pairs ($\delta_{k}$, $\delta_{\varepsilon}$) is avoided. By applying the elbow method to calibrate the values of $I_U$ for each set of independent predictions, it is found that $I_U=100$ corresponds to a reasonable approximation as well. The individual corrections obtained for $\delta_k$ and $\delta_{\varepsilon}$ can be found in Fig.~\ref{Fig_DeltaIu100_DeltaKIe0}. While both distributions appear similar, it can be proven that the corrections for $\delta_{\varepsilon}$ tend to present sharper transitions in the buffer layer near $y^{\star} = 30$ than their counterparts given by $\delta_{k}$, especially for the DNS cases requiring the largest corrections. The previous fact is found to have a large influence on the quality of the final machine learning predictions obtained for each case, since the presence of sharp transitions for $\delta_{\varepsilon}$ tends to create larger spikes in the predictions made by NNs. As a result, it is decided to employ only $\delta_k$ corrections in the final study.

\FloatBarrier
\section{Machine learning predictions}

\begin{figure}
\centering
    \input{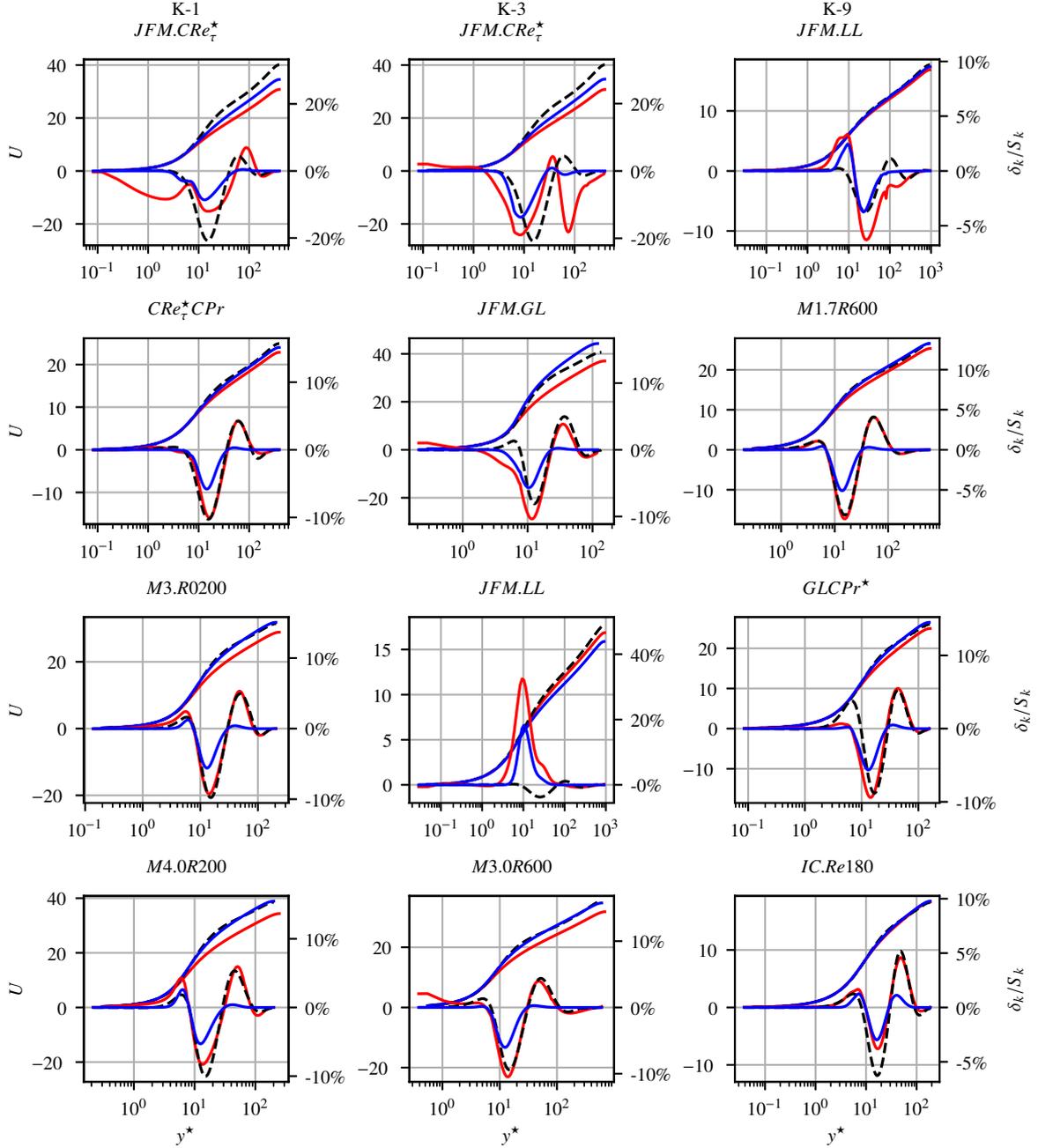}
    \centering
  	\caption{Selection of the results obtained for the K-fold validation trials at the extreme ends of the test error ranking generated during the prediction of the independent sets of ${\delta_k}$ corrections for the MK turbulence model. The upper curves represent the initial RANS velocity profiles (red dashed lines), the data-augmented velocity predictions (blue solid lines) and the reference DNS data (black dotted lines). The lower curves present the initial predictions made by the NN system (red solid lines), the results obtained after applying a weighted relaxation factor of 0.5 (blue solid lines) and the ground-truth labels for the field inversion values (black dotted lines).}
  	\label{Fig_KfoldBestDeltaK}
\end{figure}

\begin{figure}
\centering
     \input{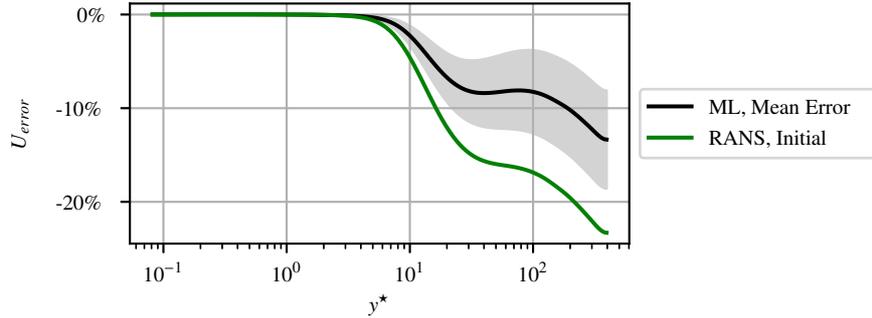}
    \centering
  	\caption{Results of the uncertainty quantification process followed for the DNS case $JFM.CRe_{\tau}^{\star}$ while employing the results of the K-fold validation runs \{1, 2, 3, 4, 7, 10\} after the prediction of the independent set of ${\delta_k}$ corrections for the MK turbulence model. The gray area corresponds to the standard deviation of the neural network predictions.}
  	\label{Fig_UncertaintyDeltaK}
\end{figure}

The final ML predictions were obtained by training the deep learning architecture described in Fig.~\ref{Fig_NNDeltaK}, and subsequently applying a weighted relaxation factor of $\alpha=0.5$ in eq. (\ref{IRF_Fixed_mag10}) to smooth all the predictions generated by neural networks. The hyper-parameter $\alpha=0.5$ was found to yield favourable results, despite substantially reducing the magnitude of the final $\delta_k$ corrections. A selection of the results can be found in Fig.~\ref{Fig_KfoldBestDeltaK}. The K-fold validation trials present in this figure correspond to the three cases with the highest modelling errors, and to the K-fold trial with the lowest error margins (K-6). The DNS cases used in the comparison also correspond to a selection of the cases with the highest and the lowest test errors found in each validation set. The results show that the ML predictions are able to retain or improve the initial predictions of the MK turbulence model in nearly every case tested. The results found for the DNS case $CRe_{\tau}^{\star}$ are analyzed in greater detail and presented Fig.~\ref{Fig_UncertaintyDeltaK}. Here, the average modelling error and corresponding standard deviation are plotted. It can be seen that the different K-fold validation trials achieve an average improvement of 13.3\%. While this improvement margin may seem modest, it is important to remember that the DNS case $JFM.CRe_{\tau}^{\star}$ corresponds to an atypical sample with respect to the remaining dataset. The extent of the modelling errors found in the DNS case $JFM.CRe_{\tau}^{\star}$ with the original MK turbulence model is unmatched by any other DNS case available. As a result, it can be seen that the ML architecture is able to achieve a robust behavior even in the presence of adverse conditions. 

Moreover, the ML system also manages to yield accurate predictions in DNS cases, which are closely related to the available training data, such as samples from the supersonic flow cases simulated by \citet{trettel2016}.
Based on the previous success, it can be concluded that the ML system is able to act as an effective non-linear interpolator, and that the predictions made for sparse DNS cases only yield moderate improvements.

Regarding the feature groups created by the NN architecture, different trends were observed across each K-fold validation run. Obtaining conclusions regarding the feature groups was complex, since the subsequent layers of hyperbolic tangent neurons in the ML system also created non-linear combinations between the different parameter groups assembled. Furthermore, the stack of features $X_f$ passed to the ML system is also likely to present complex non-linear correlations. As a result, the NN itself was able to tweak the scope and influence of each parameter group chosen during the K-fold validation runs. Despite the previous variability, it is important to note that the system achieved similar predictions across different K-fold validation trials. Reducing the size of the NN architecture was deemed unfeasible, since the ML model became incapable of fitting the training data appropriately when such attempts were made. As a result, it can only be concluded that more DNS data should be used in subsequent studies. However, it can still be noted that the ML architecture developed was able to yield positive contributions across a wide variety of unique/sparse DNS cases. Therefore, it can be concluded that the ML architecture developed is well-posed to make predictions on sparse DNS samples which may not be well-represented in the training set.


\section{Conclusions}
In this paper we used machine learning to improve the predictions of RANS turbulence modelling in channel flows subject to strong variations in their thermophysical properties. The study was based on a technique known as FIML proposed by \citet{Parish_Durai_FIML_07}. Several adaptations were introduced into the formulation proposed by the original authors, which can be recommended for future studies. The first recommendation is the use of the bold drive method with added momentum to drive field inversion optimization processes. This method proved to be unconditionally stable and numerically efficient in over 600 optimization cases run during the present work. As a result, this method can operate automatically requiring minimal attention from the user. The use of symbolic algebra solvers to generate expressions for the entries present in the matrices required by the discrete adjoint method in CFD also proved to be a valuable alternative, since the closed-form expressions generated are sparse-efficient. Furthermore, the addition of $\beta$ multipliers to specific terms in the RANS equations proved to yield numerically stable formulations. The overall shape of the corrections obtained can be controlled by employing cost functions containing adequate conversion terms (e.g., $\beta$ vs. $\delta$). Regarding this context, the use of a direct non-linear optimization approach is also recommended in the context of RANS turbulence modelling in CFD, since DNS data contains negligible levels of numerical noise or discretization errors. 

Regarding the machine learning study, the use of an initial layer of logarithmic neurons followed by layers of hyperbolic tangent neurons proved to be a robust architecture, which was able to yield positive contributions in nearly every case tested. The use of the weighted relaxation factor methodology proved to be an important asset within this context, since it was able to recover valuable trends present in otherwise spurious predictions. The high variability observed among the different K-fold validation trials did not manage to destabilize the final results. The overall behavior of the ML models indicates that the system is able to act as an excellent non-linear interpolator between DNS cases well-represented in the training set, whereas the predictions obtained for sparse DNS samples only show moderate improvement margins. Due to the previous reasons, in future studies it is recommended to search for parameter groups which display low levels of variability across different DNS cases, such that the process of building ML predictions can be simplified. The substantial variations among the K-fold validation trials also suggest that more training data should be incorporated into the study. Despite the challenges encountered, the FIML methodology developed to account for strong variations in the thermophysical properties of fluids was able to yield positive improvements across a wide variety of unique DNS cases, which were not well-represented in the training set, such as the DNS case $JFM.Re_\tau^\star$.

\bibliography{scalar}

\end{document}